\documentclass[twocolumn,aps,prx,superscriptaddress,nofootinbib]{revtex4-2}
\usepackage[T1]{fontenc}
\usepackage[utf8]{inputenc}
\usepackage{amsmath,amssymb,bm}
\usepackage{booktabs}
\usepackage{array}
\usepackage{graphicx}
\usepackage{microtype}
\usepackage[colorlinks=true,linkcolor=blue,citecolor=blue,urlcolor=blue]{hyperref}

\begin{document}

\title{Asymmetry-aided measurement-based quantum repeaters\\ and distributed quantum computing with a decoder-free client}

\author{Wooyeong Song}
\author{Sungyeon Kook}
\author{Wonhyuk Lee}
\author{IlKwon Sohn}
\email{d2estiny@kisti.re.kr}
\affiliation{Quantum Network Research Center, Korea Institute of Science and Technology Information (KISTI), Daejeon 34141, Republic of Korea}

\date{\today}

\begin{abstract}
Distributed quantum computation needs to move logical qubits across lossy
optical links, yet this transmission layer is usually designed separately from
the computation it serves. We treat the two together by recognizing that a
measurement-based quantum repeater is a two-dimensional code \emph{foliated}
along the transmission axis, so that the dominant channel loss is concentrated
on the transmitted sector while the locally measured qubits are largely spared.
Matching a code's distance to this structural asymmetry, we show that a rectangular Bacon--Shor subsystem code transmits a logical qubit markedly more efficiently than transmission-unaware encodings. Over continental distances, its cost-optimal repeater density is about an order of magnitude lower than that of a recent $[[48,6,8]]$ benchmark at comparable transmission rate, and roughly half that of a symmetric code of equal size. Moreover, we extend the framework to a central-to-client round trip in which a code-level, distance-preserving code switch joins the transmission legs to the client's computation, and joint decoding of the heterogeneous syndrome record at the central node lets distributed quantum computation proceed with a \emph{decoder-free} client.
\end{abstract}

\maketitle

\section{Introduction}

% ============================================================
% Introduction
% ============================================================
Distributed quantum computation (DQC) spreads a single computation across several
quantum processors linked by quantum channels~\cite{cirac1999}, extending the reach of any one
device beyond the qubits it can hold locally and ultimately toward a future quantum internet~\cite{kimble2008,wehner2018}. Because entanglement between distant
nodes cannot be created by local operations and classical communication, it must be
physically distributed across the network, so a loss-tolerant quantum-transmission
layer is indispensable, not incidental, to DQC. The bulk of work on DQC concentrates
on the computation itself (distributed gate protocols~\cite{eisert2000}, gate teleportation~\cite{gottesman1999}, modular processor architectures~\cite{monroe2014}, and
the distribution of entanglement) and treats the delivery of an encoded
logical qubit across the network as a given primitive. Transmitting a logical
qubit over a lossy optical channel is, however, a decisive obstacle in its own
right. Photon loss attenuates a fiber link exponentially in its length, and
without a loss-tolerant way to move encoded information between nodes the
distributed computation never assembles. Recent work has studied fault-tolerant distributed quantum computing and its network-level entanglement overheads~\cite{nickerson2013,ramette2024,chandra2025,naito2026}, while one-way repeaters encode the transmitted qubit against loss as a long-distance communication primitive~\cite{jiang2009,muralidharan2016,wo2023}. A unified code-design perspective that treats logical-state transmission and logical computation within the same measurement-based repeater architecture, however, remains comparatively underdeveloped. We argue that this transmission layer
deserves to be treated on the same footing as the computation it serves, and we
show that doing so changes how a distributed computation can be organized.

\begin{figure*}[t]
\centering
\includegraphics[width=\linewidth]{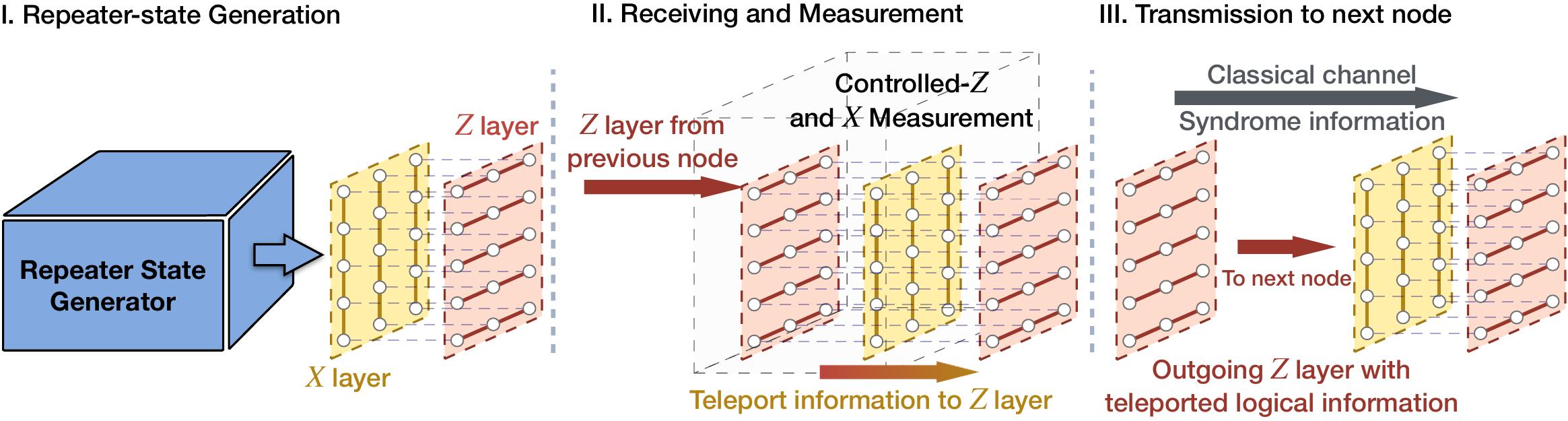}
\caption{The measurement-based repeater as a foliated code (stages I--III, left
to right). (I)~A repeater-state generator (RSG) prepares two 2D code layers, a
local $X$ layer and a transmitted $Z$ layer, linked by qubitwise CZ bonds.
(II)~The incoming $Z$ layer from the previous node is stitched to the local $X$
layer by qubitwise CZ bonds. The enclosed two-layer block undergoes the
controlled-$Z$ and $X$ measurement, teleporting the logical information onto the
outgoing $Z$ layer. (III)~The outgoing $Z$ layer is transmitted to the next node,
while the measurement outcomes are forwarded as syndrome data over a classical
channel. Dashed lines denote qubitwise CZ bonds. Ancilla qubits are omitted for
clarity. Consistent with the asymmetric Bacon--Shor convention, each transmitted $Z$
layer is drawn as $m$ short gauge chains of length $n$, while each local $X$
layer is drawn as $n$ longer chains of length $m$. This depicts the local
foliated-layer connectivity. The corresponding dressed-logical distances (the minimum weight over
gauge-equivalent representatives) are $(d_Z,d_X)=(m,n)$, so the channel-exposed sector receives the larger distance.
The transmitted $Z$ layers cross the fiber and suffer the channel erasure rate
$p_Z$, whereas the locally prepared $X$ layers suffer only $p_X<p_Z$.}
\label{fig:foliation}
\end{figure*}

Preserving high-fidelity entanglement across a lossy link is the task of a quantum repeater, needed not only over continental distances but on any link where photon loss would degrade the shared state, including the shorter links within a distributed processor. Conventional repeaters store the state in matter quantum memories~\cite{briegel1998} and
build up entanglement over many two-way signaling rounds, demanding long coherence
times and added latency. All-photonic
repeaters~\cite{azuma2015,pant2017,borregaard2020,azuma2023} remove the memory
entirely, encoding the link in a large entangled photonic resource state operated
one-way---a fast, memoryless transmission layer well matched to a photonic
distributed-computing platform. The measurement-based repeater of Niu
\emph{et al.}~\cite{niu2023} is one such instance. A logical qubit encoded in a
Calderbank--Shor--Steane (CSS) code is prepared as a photonic graph state,
propagated node to node, and measured, so that the encoded state is teleported one
step downstream at each station. Because photon loss is heralded, every lost
photon is an erasure at a known location, and the reach and rate of the
repeater are set by how efficiently its code tolerates erasure. The codes used for
this purpose, most commonly the surface code~\cite{fowler2012}, protect every physical qubit
equally. Yet the repeater channel is not symmetric. Only the qubits that actually
cross the fiber are exposed to the channel loss, while those prepared and measured
inside a station are not. This structural asymmetry is intrinsic to the
protocol but invisible to a \emph{transmission-unaware} encoding, and leveraging it is the
opportunity we pursue. The lesson is structural. The transmission stage should carry
its own code, fitted to the channel, rather than the computation's code
stretched across the link.

We make the asymmetry explicit by viewing the repeater as a \emph{foliated}
two-dimensional code whose foliation axis is the transmission direction, and we
exploit it with a code whose protection is deliberately lopsided. It is a rectangular
Bacon--Shor subsystem code with a large distance in the channel-exposed sector and
a small one locally. Because its erasure correction reduces to that of repetition
codes, its loss threshold attains the maximal value of $100\%$, so that, for the sector-selective heralded photon loss considered here, there is no percolation threshold to cross and enlarging
the code keeps suppressing the failure at any loss below unity. This does not beat
the $50\%$ no-cloning bound for an unknown qubit on an erasure
channel~\cite{bennett1997}. It is the loss asymmetry, not a symmetric code,
that we exploit. It transmits a logical qubit at a markedly higher effective rate
than a symmetric code of equal size, both per link and over continental distances. We then observe that the repeater protocol is itself a
measurement-based quantum error correction process, which lets us interpret
transmission and computation within a single framework spanning an outbound transmission, a
computation at the edge, and a return transmission, decoded together. In a DQC
setting this has a concrete payoff. The entire heterogeneous syndrome record is
decoded once, jointly, at the central node, so that the client computes with no
global decoder of its own, while the Bacon--Shor/surface-code switch that joins the
two encodings remains distance-preserving at the code level. The remainder of this paper develops these two results in turn. These are the asymmetric-code transmission advantage (Secs.~\ref{sec:asymcode}--\ref{sec:longdist}) and the deferred-decoding distributed computation it enables (Secs.~\ref{sec:switch} and~\ref{sec:deferred}).

% ============================================================
% Measurement-based repeaters as foliated codes
% ============================================================
\section{Measurement-based repeaters as foliated codes}
Concretely, this teleportation along the chain (Alice $\to$ repeaters $\to$ Bob)
is realized station by station. A fresh code block is prepared as a photonic graph
state, joined to the incoming block by a transversal controlled-phase operation, and measured together with it, transferring the encoded state onto the outgoing block.

It is useful to view this process as a \emph{foliated} quantum
code~\cite{bolt2016}. Repeating and measuring a two-dimensional CSS code round by
round generates a three-dimensional cluster state, the extra dimension being the foliation axis. For the repeater this axis is precisely the transmission direction. A 2D code block is foliated into a 3D cluster threaded
along the fiber (Fig.~\ref{fig:foliation}). Each slice $\ell$ is a copy of the 2D
code, and consecutive slices are linked so that measuring one transfers the
logical operators to the next. The CSS structure splits the foliation into alternating primal ($Z$) and dual ($X$) layers, carrying logical-$Z$ and logical-$X$ information respectively, at slices $\ell=\dots,2k{-}1,\,2k,\,2k{+}1,\dots$ (Fig.~\ref{fig:foliation}).

What makes this picture more than a relabeling is that the layers are not
physically equivalent. The data qubits of the transmitted layers traverse the
lossy fiber between stations, whereas the qubits prepared and measured locally
within a station never enter the channel. This distinction is invisible in a bare
2D code but explicit once the code is foliated along the transmission axis, and
it is the origin of the asymmetry we exploit below.
% ============================================================
% Asymmetric channel, asymmetric code
% ============================================================
\section{Asymmetric channel, asymmetric code}
\label{sec:asymcode}
We now quantify this structural asymmetry. With the standard two-parameter loss
model~\cite{niu2023}, the transmitted and local layers incur erasure at the
distinct rates
\begin{equation}
 p_Z = 1-\eta_r(1-p_{\rm ch}),\qquad
 p_X = 1-\sqrt{\eta_r},
 \label{eq:rates}
\end{equation}
where $\eta_r$ is the repeater efficiency (or transmittance), collecting the photon-source and detector efficiency, on-chip loss, and in/out coupling losses, $p_{\rm ch}$ the per-segment channel
loss with $1-p_{\rm ch}=10^{-\alpha_0 L/10}$ the fiber transmission over a length $L$ ($\alpha_0=0.2\,{\rm dB/km}$). Here $L_{\rm att}=10/(\alpha_0\ln 10)\simeq21.7$~km is the corresponding fiber attenuation length. Equivalently, channel-traversing qubits
transmit with probability $\eta_r(1-p_{\rm ch})$ and internal (locally prepared)
qubits with $\sqrt{\eta_r}$, reproducing the model of Ref.~\cite{niu2023}. The two
rates differ markedly. The transmitted-layer erasure rate $p_Z$ is set by the link and grows
with distance, whereas the local rate $p_X$ is fixed by the node hardware alone.
Hence $p_Z>p_X$ for any nonideal repeater node, and, crucially, the disparity
widens as the nodes are placed farther apart. Following Ref.~\cite{niu2023}, all
syndrome (ancilla) qubits are generated and measured locally and therefore
experience only the local erasure rate $p_X$. We adopt this convention throughout the main text so
that every code is compared on the same footing as the surface code. The more
conservative variant, in which the syndromes of transmitted layers additionally
suffer the transmitted erasure rate $p_Z$, is reported in Appendix~\ref{app:asis} and leaves
all qualitative conclusions intact.

The labels $X$ and $Z$ in Eq.~\eqref{eq:rates} refer to the two CSS/foliated
sectors of the construction, not to physical Pauli $X$- or $Z$-error channels.
Throughout we reserve loss for the physical channel and node parameters
($p_{\rm ch}$ and $\eta_r$) and erasure rate for the per-qubit
probabilities $p_Z,p_X$ seen by the two sectors, since heralded photon loss
enters the code as erasure at known locations. The heralding assumed here is the
standard photonic one. A successful measurement yields a definite detector click,
so a no-click event flags the loss at a known location, as for dual-rail or
polarization qubits. It fails, and a lost photon instead enters as an
\emph{unheralded} Pauli error, under single-rail encodings (where an absent photon
is itself a valid outcome), dark counts, or non-number-resolving detection, and
erasure-conversion techniques~\cite{wu2022} are designed to keep the dominant
losses heralded. A
CSS code protects the two sectors independently, with distances $d_X$ and $d_Z$.
We write $P_L^X$ and $P_L^Z$ for the corresponding logical failure probabilities
and
\[
 P_L=1-(1-P_L^X)(1-P_L^Z)\approx P_L^X+P_L^Z
\]
for the total. Here the $Z$-layer is exposed to the large erasure rate $p_Z$, and
the $X$-layer only to $p_X$. A code with $d_X=d_Z$, symmetric between its two sectors as in a square surface-code patch or in small symmetric CSS codes, thus spends its qubits equally on two unequal threats, overprotecting the
benign sector and underprotecting the dominant one. The natural remedy is an
asymmetric code, $d_Z>d_X$, placing protection where the loss concentrates.

Although erasures and Pauli errors are decoded differently, the same
distance-allocation logic applies at the level of sector failure probabilities.
In both cases the below-threshold logical failure rate in a sector is suppressed
exponentially in the corresponding distance. For erasures the locations are
known, so the relevant failure event is that the erased set supports a nontrivial
dressed logical operator~\cite{grassl1997}. A useful below-threshold estimate is therefore
\begin{equation}
 P_L^\sigma \sim
 \left(\frac{p_\sigma}{p_{\rm th}}\right)^{c\,d_\sigma},
 \qquad
 \sigma\in\{X,Z\},
 \label{eq:scaling}
\end{equation}
with an order-unity constant $c$ and erasure threshold $p_{\rm th}$. Balancing the
two sector contributions at fixed budget gives the heuristic condition
\begin{equation}
 \frac{d_Z}{d_X}
 \sim
 \frac{\ln(p_{\rm th}/p_X)}{\ln(p_{\rm th}/p_Z)}
 >1.
 \label{eq:balance}
\end{equation}
The optimal code is therefore expected to be elongated along the transmitted,
high-loss sector, increasingly so as the link grows lossier. Equation~\eqref{eq:balance}
is not an exact finite-size optimum. It captures the trend that the distance
allocation should track the architectural erasure asymmetry. The argument invokes no
particular code. Any CSS family with independently tunable $(d_X,d_Z)$ inherits
the advantage.

The Bacon--Shor (BS) code~\cite{bacon2006} is the most transparent realization. Its two distances are
fixed directly by the lattice dimensions. With the convention
\[
 (d_Z,d_X)=(m,n),
\]
the aspect ratio $m:n$ sets the code asymmetry, with the square $m=n$ recovering
the symmetric case. This parallels the Bacon--Shor
response to biased Pauli noise, where an asymmetric block exploits a
strong Pauli bias by tuning its aspect ratio~\cite{napp2013,brooks2013}. We use that result
only as a design guide. Exploiting a loss (erasure) bias in this way is the
heralded-loss analogue of biased-erasure error correction~\cite{wu2022,sahay2023prx}.
While Shor-type codes have been demonstrated for loss tolerance in all-photonic
repeaters~\cite{zhang2022}, that was done in a different (repeater-graph-state) architecture, with a symmetric code, and for transmission only, whereas we match an asymmetric code to the foliation and extend it to a
computation round trip. In the present setting the bias is not an intrinsic
Pauli-noise bias of the hardware but a structural erasure bias generated
by the measurement-based architecture. The shared lesson is that when the
physical process treats the two CSS sectors unequally, the code distances should
not be allocated symmetrically.

The short chains in Fig.~\ref{fig:foliation} represent the local gauge-chain
connectivity of each foliated layer. The erasure distance is determined by the
minimum dressed logical support. For the Bacon--Shor code this principle has a
simple erasure interpretation. The leading erasure mechanism is controlled by the
minimum dressed-logical weight in each sector. The leading $Z$-sector logical erasure events are erased sets
supporting a length-$m$ $Z$-type dressed-logical representative, and the leading
$X$-sector events support a length-$n$ representative.
Thus, below threshold,
\begin{equation}
 P_L^Z \sim A_Z(m,n)\,p_Z^{\,m},
 \qquad
 P_L^X \sim A_X(m,n)\,p_X^{\,n},
 \label{eq:detour}
\end{equation}
where the prefactors $A_Z,A_X$ depend on finite-size geometry and gauge
degeneracy but remain subexponential in the distances. The important point is the
exponent. Assigning the larger dimension $m$ to the channel-exposed $Z$ sector
suppresses its frequent erasures exponentially, while the locally generated $X$
sector can be assigned the smaller dimension $n$. The Bacon--Shor aspect ratio
therefore converts the structural erasure bias into a distance allocation matched
to the two erasure rates. Equivalently, increasing $m$ increases the minimum erased
support needed to realize a $Z$-sector logical failure. It should not be
interpreted as providing surface-code-like local detours around erasure
holes (Appendix~\ref{app:gauge}).

We confirm the trend numerically. Holding the qubit count fixed at
$N=d_Zd_X=45$ and varying only the aspect ratio (Table~\ref{tab:aspect}), the
logical failure probability is large for both near-square and extreme shapes and
is minimized by an elongated code whose optimal aspect ratio grows with loss. The optimum is interior, the shorter sector first securing enough distance to cover its own loss before the remaining budget lengthens the channel-exposed sector. At $p_{\rm ch}=0.1$ the minimum sits near $d_Z/d_X\simeq1.8$ ($9\times5$),
shifting to $d_Z/d_X=5$ ($15\times3$) by $p_{\rm ch}=0.2$. This is exactly what
Eq.~\eqref{eq:balance} anticipates. The optimal asymmetry tracks the channel
asymmetry, the lossier the link the more elongated the favored code.

\begin{figure}[t]
\centering
\includegraphics[width=\columnwidth]{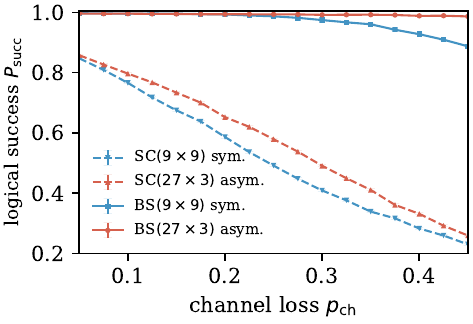}
\caption{Logical success probability of four equal-size ($N=81$) codes versus
channel loss ($\eta_r=0.9$, one repeater), all decoded with the same
maximum-likelihood foliated erasure decoder. The asymmetric Bacon--Shor
code $\mathrm{BS}(27\times3)$ sustains the highest $P_{\rm succ}$ across the whole loss range,
whereas the same elongation improves the surface code only modestly
($\mathrm{SC}(9\times9)\to\mathrm{SC}(27\times3)$). Matching the aspect ratio to the
channel bias pays off for the gauge-rich subsystem code but not for the topological
one. Each point is $2\times10^4$ Monte Carlo erasure samples. Error bars are
$\pm1$ standard error and are smaller than the markers (maximum SE $\approx0.004$).}
\label{fig:n81}
\end{figure}

\begin{figure*}[t]
\centering
\includegraphics[width=0.92\linewidth]{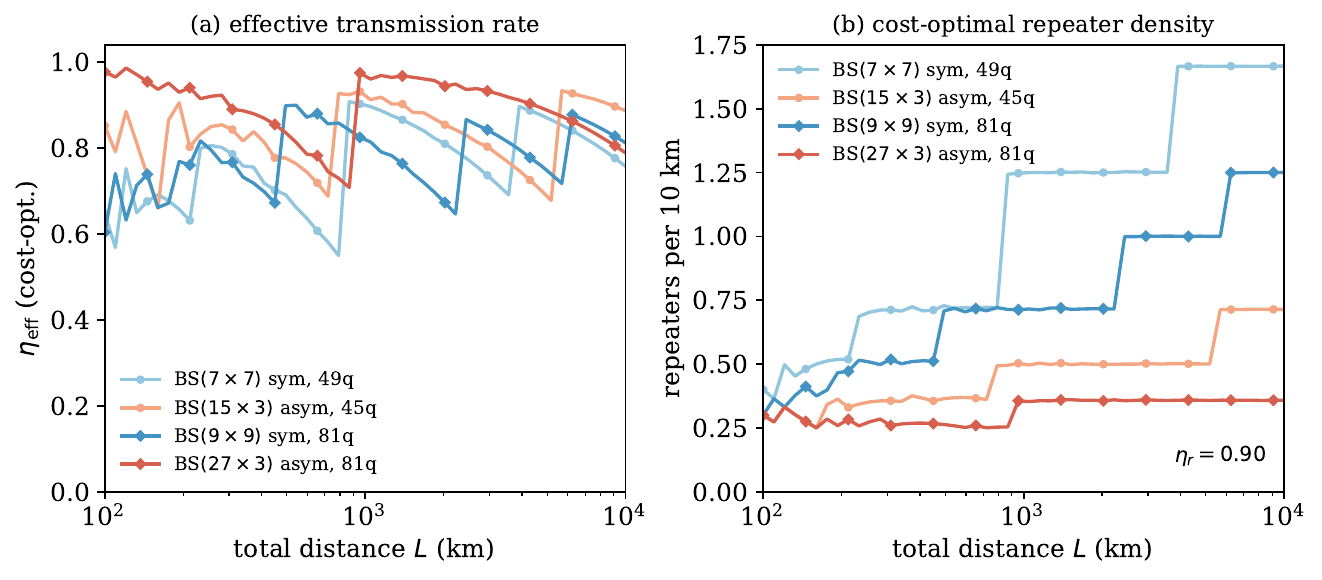}
\caption{\textbf{Performance of optimized repeater architectures.}
Long-distance performance at repeater efficiency $\eta_r=0.9$ (annotated in panel
(b), lossy local ancilla $p_X=1-\sqrt{\eta_r}\approx0.05$, reliable
transmitted-sector syndrome in the well-scheduled cross-layer regime, with the other
efficiencies $\eta_r=0.85,0.95$ in Fig.~\ref{fig:longdist-app}). Four
equal-pair codes are compared. These are the $\sim\!48$-qubit pair
$\mathrm{BS}(7\times7)$ (symmetric) and $\mathrm{BS}(15\times3)$ (asymmetric), and
the $81$-qubit pair $\mathrm{BS}(9\times9)$ (symmetric) and
$\mathrm{BS}(27\times3)$ (asymmetric). (a)~Cost-optimal effective transmission
rate $\eta_{\rm eff}$ versus total distance $L$. (b)~Corresponding cost-optimal repeater density (repeaters per $10$~km). At matched code size the asymmetric codes
(warm colors) sit at the higher $\eta_{\rm eff}$ in (a) and the lower repeater
density in (b), roughly half the stations of their symmetric counterparts (cool
colors), across the whole range. The discontinuities are the integer steps of
the discrete repeater-number optimization, as in Ref.~\cite{niu2023}. Per-hop
rates are Monte-Carlo erasure simulations decoded with the maximum-likelihood criterion in the reliable-syndrome limit (syndrome availability set
to unity). The cascade follows Eq.~\eqref{eq:cascade}. This is the effective
foliated model discussed in the text, not a full circuit-level teleportation
chain.}
\label{fig:longdist}
\end{figure*}

\begin{figure*}[t]
\centering
\includegraphics[width=0.92\linewidth]{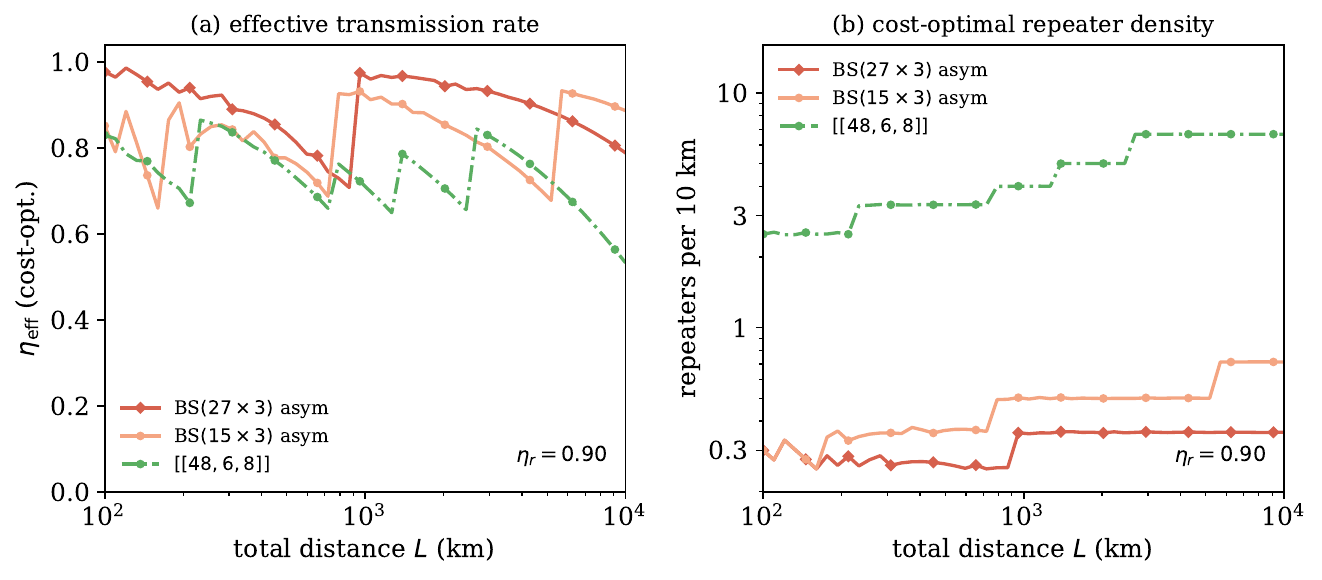}
\caption{\textbf{Single common-simulation comparison with previous work.}
(a)~Cost-optimal effective transmission rate $\eta_{\rm eff}$ and (b)~cost-optimal repeater density (logarithmic axis) versus total distance $L$ at $\eta_r=0.9$, for the asymmetric
$\mathrm{BS}(27\times3)$ (81 data qubits) and $\mathrm{BS}(15\times3)$ (45 qubits) and
the $[[48,6,8]]$ generalized-bicycle code of Ref.~\cite{niu2023} (48 qubits), all
decoded in one common simulation with the same optimal maximum-likelihood erasure
decoder (at least as strong as the Algorithm~1 of Ref.~\cite{niu2023}). The asymmetric
Bacon--Shor codes are cost-optimal at $\sim\!0.3$--$0.5$ stations per $10$~km, an order of
magnitude below the $\sim\!3$--$7$ per $10$~km of $[[48,6,8]]$, at comparable or higher
$\eta_{\rm eff}$. The $[[48,6,8]]$ curve reproduces the metrics reported in
Ref.~\cite{niu2023} (Table~\ref{tab:niu-compare}). $\mathrm{BS}(15\times3)$ is
resource-matched to the benchmark. Steps are the integer repeater-number optimization.}
\label{fig:niu-common}
\end{figure*}

\begin{table}[t]
\caption{Fixed-budget aspect-ratio scan of the Bacon--Shor transmission code at
$N=d_Zd_X=45$ ($\eta_r=0.9$, one repeater, $3\times10^4$ samples, with odd dimensions
used in this transmission-only scan, for consistency with the finite-size
simulation layouts).
The sector contributions $P_L^Z$, $P_L^X$ are shown at $p_{\rm ch}=0.1$. The last
column gives the total $P_L$ at $p_{\rm ch}=0.1$ and $0.2$.}
\label{tab:aspect}
\begin{ruledtabular}
\begin{tabular}{cccccc}
$d_Z$ & $d_X$ & $d_Z/d_X$ & $P_L^Z$ & $P_L^X$ & $P_L\;(0.1\mid0.2)$\\
\hline
1  & 45 & 0.02 & 0.408  & 0.0005 & $0.409\mid0.533$\\
3  & 15 & 0.20 & 0.070  & 0.0005 & $0.071\mid0.157$\\
5  & 9  & 0.56 & 0.015  & 0.0008 & $0.016\mid0.046$\\
9  & 5  & 1.80 & 0.0027 & 0.0004 & $\mathbf{0.003}\mid0.007$\\
15 & 3  & 5.00 & 0.0025 & 0.0013 & $0.004\mid\mathbf{0.005}$\\
45 & 1  & 45.0 & 0.0029 & 0.102  & $0.105\mid0.102$\\
\end{tabular}
\end{ruledtabular}
\end{table}

% ============================================================
% Asymmetric Bacon--Shor improves the repeater
% ============================================================
\section{Asymmetric Bacon--Shor improves the repeater}
\label{sec:perhop}
Having established the principle, we apply it to the proposed repeater and
measure the logical success probability $P_{\rm succ}=(1-P_L^X)(1-P_L^Z)$ versus
the channel loss $p_{\rm ch}$, at fixed qubit budget $N=d_Zd_X$ with one repeater ($\eta_r=0.9$).

Figure~\ref{fig:n81} compares four equal-size ($N=81$) data-code blocks. Here
$N$ denotes the number of data qubits in the two-dimensional code block. A full
photonic-resource comparison including syndrome and graph-state ancillas can be
added separately. All four blocks are decoded on a common footing, the same
foliated primal/dual erasure model with locally measured syndromes and the
optimal (maximum-likelihood) erasure decoder~\cite{delfosse2020}. A symmetric surface-code (SC) patch $\mathrm{SC}(9\times9)$ performs poorly as a transmission block in this
architecture. Its $Z$-sector failure is dominated by the accumulated
transmitted-layer erasures, so its $P_{\rm succ}$ degrades steadily with channel loss, from
$0.85$ at $p_{\rm ch}=0.05$ to $0.23$ at $p_{\rm ch}=0.45$. An independent
direct-erasure benchmark reproduces the expected $\sim\!50\%$ surface-code
threshold (Appendix~\ref{app:gauge}), confirming that the poorer transmission
performance reflects the code structure rather than the decoder. Crucially,
asymmetry does not rescue the surface code. Elongating
it to $\mathrm{SC}(27\times3)$ improves $P_{\rm succ}$ only modestly. The gain
reaches $\approx0.08$ at intermediate loss but shrinks again toward heavy loss
($0.26$ versus $0.23$ at $p_{\rm ch}=0.45$), leaving the surface code a poor
transmission code (Fig.~\ref{fig:n81}). The reason is
the erasure threshold. The transmitted $Z$-sector already operates at or
above the surface code's $\sim\!50\%$ erasure threshold~\cite{stace2009}, and above threshold a
larger $d_Z$ buys no sub-threshold distance suppression. The logical-failure
rate stays $O(1)$ however elongated the patch. It is precisely the Bacon--Shor
gauge freedom that changes this. By supplying many equivalent dressed logical
representatives it lowers the effective erasure rate seen by the $Z$
logical, keeping that sector below threshold, so that enlarging $d_Z$ then
suppresses $P_L^Z$ as $\sim p_Z^{\,d_Z}$. Matching the code asymmetry to the
channel bias therefore pays off for the subsystem code but not for the
topological one. Both Bacon--Shor codes do far better,
but they degrade differently. The symmetric $\mathrm{BS}(9\times9)$ sustains
$P_{\rm succ}>0.97$ up to $p_{\rm ch}\simeq0.3$ and then falls to $0.89$ at
$p_{\rm ch}=0.45$, whereas the asymmetric $\mathrm{BS}(27\times3)$, concentrating
distance in the transmitted sector, maintains $P_{\rm succ}\ge0.98$ throughout.

It is worth separating two distinct effects in Fig.~\ref{fig:n81}. The gap $\mathrm{SC}(9\times9)\!\to\!\mathrm{BS}(9\times9)$ is a code-family effect. The further gain $\mathrm{BS}(9\times9)\!\to\!\mathrm{BS}(27\times3)$ is the aspect-ratio effect. The former is by far the larger, so the gauge structure, not the asymmetry per se, is the dominant ingredient, and the asymmetry is the secondary optimization it enables. A row-space diagnostic in Appendix~\ref{app:gauge} isolates this gauge-structure origin and confirms the repetition-code scaling $P_L^Z\sim p_Z^{\,m}$ [Eq.~\eqref{eq:detour}] directly at erasure rates at and above the surface code's $\sim\!50\%$ threshold, where the surface-code logical stays saturated and flat in distance (Fig.~\ref{fig:gauge}).

Consistent with Eq.~\eqref{eq:balance}, the asymmetric code is not uniformly
superior. At low loss the symmetric code is marginally better, since the benign
$X$ sector is not yet starved by shrinking $d_X$. The asymmetric advantage sets
in once $p_{\rm ch}\gtrsim0.15$ and grows thereafter, reaching a factor
$\sim\!10$ in $P_L^Z$ at $p_{\rm ch}=0.45$ ($0.011$ versus $0.114$). The optimal
aspect ratio thus tracks the channel loss (the lossier the link, the more
elongated the favored code), exactly the behavior of the fixed-budget scan in
Table~\ref{tab:aspect}. The elongation is nonetheless bounded from above. The
practical limit set by the quality of syndrome extraction is discussed below.

In short, matching the code asymmetry to the intrinsic channel asymmetry of the
repeater turns a finite qubit budget into a transmission code that remains
near-deterministic ($P_{\rm succ}\gtrsim0.98$) even at $45\%$ per-segment loss,
where a symmetric allocation of the same resources has already degraded
substantially.

The asymmetric advantage belongs to the error-corrected repeated link rather
than to a bare point-to-point hop: an intermediate repeater must re-extract the
transmitted-sector syndrome for the $p_Z^{d_Z}$ suppression to be realized. The
optimal asymmetry is therefore finite, bounded below by starvation of the local
sector ($d_X=1$ fails outright in Table~\ref{tab:aspect}) and above by the quality
of syndrome extraction along the elongated sector. Although a one-shot syndrome
can over-penalize an elongated block, cross-layer redundancy restores the
reliable-syndrome limit after only two or three foliation layers
(Appendix~\ref{app:crosslayer}). We use this regime in the multi-hop analysis
below.

\section{Long-distance performance and repeater spacing}
\label{sec:longdist}
The preceding section concerns a single error-corrected hop. To assess the architecture over
continental distances we cascade the foliated link, following the resource model
of Ref.~\cite{niu2023}. For a chain divided into $N$ equal elementary links across a total distance $L$, with $N-1$ intermediate repeater stations, the per-hop spacing is $L_0=L/N$ and, because each hop teleports the
logical state independently, the end-to-end effective transmission rate
factorizes as
\begin{equation}
 \eta_{\rm eff}(L_0,L)=P_{{\rm succ},1}(L_0)^{N},\qquad N=L/L_0,
 \label{eq:cascade}
\end{equation}
with $P_{{\rm succ},1}$ the single-hop rate of Sec.~\ref{sec:perhop} evaluated at the
per-segment loss $1-p_{\rm ch}=10^{-\alpha_0 L_0/10}$. Equation~\eqref{eq:cascade}
is exponential in $L$, matching the form fitted in Ref.~\cite{niu2023}. For each
code and total distance we place repeaters at the cost-optimal spacing, minimizing
the resource-to-performance ratio $C=(N/L)/\eta_{\rm eff}$. This is the cost
function of Ref.~\cite{niu2023} with its code-size factor $n_{\rm phys}/n_{\rm
log}$, which is constant for a fixed code, dropped from the per-code optimization.

Gauge fixing maps the foliated Bacon--Shor code onto a pair of repetition-code
foliations, one per sector~\cite{brown2020,pesah2025}, so the per-hop construction
stacks along the transmission axis as a CSS code does. A suitable gauge schedule
keeps the spacetime detectors at constant weight and recovers a
threshold~\cite{alam2025}. We therefore use an effective foliated model in which
each hop is projected onto the reliable-syndrome erasure correctability of
Sec.~\ref{sec:perhop}, allowing the single-hop scaling $P_L^Z\sim p_Z^{d_Z}$ to
carry over through Eq.~\eqref{eq:cascade}. Appendix~\ref{app:crosslayer} shows that
this limit is reached at cross-layer depth $T=2$--$3$, while the conservative
channel-exposed-syndrome model preserves the same ordering
(Table~\ref{tab:asis}). As in Ref.~\cite{niu2023}, we model heralded loss exactly
and take the entangling operations that build the foliation to be ideal; we do
not simulate the full circuit-level teleportation chain.

Figure~\ref{fig:longdist} shows the outcome at $\eta_r=0.9$ (companion plots at
$\eta_r=0.95,\,0.85$ in Appendix~\ref{app:longdist}). Every Bacon--Shor code keeps $\eta_{\rm eff}$ far above direct fiber transmission ($10^{-\alpha_0 L/10}$, already $10^{-2}$ at $L=10^2$~km) across $10^2$--$10^4$~km, but at sharply different resource cost. The asymmetric codes are cost-optimal at a
much sparser repeater spacing than the symmetric ones of equal size. At
$L=10^3$~km the asymmetric $\mathrm{BS}(27\times3)$ places repeaters at
$L_0\simeq1.3\,L_{\rm att}$ (one per $\sim29$~km) at $\eta_{\rm eff}\simeq0.95$,
whereas the symmetric $\mathrm{BS}(9\times9)$ requires $L_0\simeq0.6\,L_{\rm att}$
(one per $\sim14$~km) and reaches only $\eta_{\rm eff}\simeq0.85$, roughly half
the stations at a higher transmission rate. The same ordering holds at the
$\sim\!48$-qubit scale of compact quantum-low-density-parity-check blocks. The
asymmetric $\mathrm{BS}(15\times3)$ needs about half the repeaters of the
symmetric $\mathrm{BS}(7\times7)$. Thus the asymmetric advantage established per-hop in Sec.~\ref{sec:perhop} compounds over distance into a markedly lower repeater
density, the resource that dominates the cost of a long link.
Table~\ref{tab:niu-compare} places our metrics, in the format of
Ref.~\cite{niu2023}, beside the $[[48,6,8]]$ generalized-bicycle benchmark
reported there. The comparison is drawn under matched loss-model assumptions and in the same reliable-syndrome regime. Both sides use the same two-parameter loss model [Eq.~\eqref{eq:rates}], the same repeater efficiencies $\eta_r=0.85,0.9,0.95$, and the same cost function, with its code-size factor held fixed for the per-code spacing optimization. Syndromes are locally generated and taken effectively reliable through joint/cross-layer decoding. The $[[48,6,8]]$ figures are the literature benchmark of
Ref.~\cite{niu2023}. The Bacon--Shor figures are our maximum-likelihood erasure simulations, whose correctability test reproduces the analytic
$[[7,1,3]]$ result of that reference. We additionally evaluate $[[48,6,8]]$ within our own framework under the
identical maximum-likelihood erasure decoder (Fig.~\ref{fig:niu-common}), reproducing
its reported metrics and placing both codes in a single common simulation. On this basis
the asymmetric Bacon--Shor codes are cost-optimal at a repeater spacing
$L_0/L_{\rm att}$ an order of magnitude larger than the $[[48,6,8]]$ benchmark
($\sim\!130\%$ versus $4$--$14\%$), at comparable or higher $\eta_{\rm eff}$.

This gap reflects a structural difference rather than an artifact of the
comparison. The surface code and the $[[48,6,8]]$ block are governed by a
percolation-type erasure threshold near $50\%$, so to keep $\eta_{\rm eff}$ high
their cost optimum lies well below threshold, at small $L_0$ and hence dense
stations. In the effective sector model used here the Bacon--Shor transmitted
sector instead fails through a repetition-like distance scaling $P_L^Z\sim
p_Z^{d_Z}$ [Eq.~\eqref{eq:detour}] rather than through a two-dimensional
percolation threshold, so a large $d_Z$ keeps the per-hop failure small even at
$p_Z\approx0.8$ (a per-segment loss of $\sim\!80\%$, i.e.\ $L_0$ of order
$L_{\rm att}$), where the threshold-limited codes have already failed. The
advantage thus traces to the code structure (the Bacon--Shor gauge freedom)
rather than to the reliable-syndrome assumption itself. Granting the competing
codes ideal syndromes would not move their percolation threshold, and hence not
their dense optimum.

The price is paid in physical qubits per logical qubit. A single-logical-qubit ($k=1$) Bacon--Shor block spends $n/k=45$--$81$ physical qubits per logical qubit against $n/k=8$ for the $[[48,6,8]]$ code (Table~\ref{tab:niu-compare}). The asymmetric construction trades
physical-qubit overhead at each station for a much lower density of stations.
This is the favorable trade whenever the deployed repeater stations, each a
full photonic apparatus, dominate the cost over the qubits within a station,
and it is the regime in which the asymmetric Bacon--Shor repeater is most
attractive.
\begin{table}[t]
\caption{Long-distance repeater metrics for the foliated Bacon--Shor codes at
$\eta_r=0.9$, in the format of Ref.~\cite{niu2023}, alongside the $[[48,6,8]]$
generalized-bicycle code evaluated there under the same loss model. Ranges span
the long-haul regime $L=10^3$--$10^4$~km. The asymmetric codes (large $d_Z$,
small $d_X$) are cost-optimal at a far larger repeater spacing $L_0/L_{\rm att}$
(sparser stations) while sustaining comparable or higher $\eta_{\rm eff}$. The
trade-off is a larger qubit-per-logical ratio $n/k$ (here every Bacon--Shor block
has $k=1$). Effective rates are maximum-likelihood erasure Monte Carlo
($\eta_{\rm eff}$ carries a few-percent statistical uncertainty).}
\label{tab:niu-compare}
\begin{ruledtabular}
\begin{tabular}{lcccc}
code & $n/k$ & $(d_Z,d_X)$ & $\eta_{\rm eff}$ & $L_0/L_{\rm att}$\\
\hline
$\mathrm{BS}(7\times7)$ (sym.)   & 49 & $(7,7)$  & 0.70--0.90 & 28--37\%\\
$\mathrm{BS}(15\times3)$ (asym.) & 45 & $(15,3)$ & 0.66--0.93 & 64--92\%\\
$\mathrm{BS}(9\times9)$ (sym.)   & 81 & $(9,9)$  & 0.63--0.88 & 37--64\%\\
$\mathrm{BS}(27\times3)$ (asym.) & 81 & $(27,3)$ & 0.79--0.97 & \textbf{128--132\%}\\
\hline
$[[48,6,8]]$~\cite{niu2023}      & 8  & $(8,8)$  & 0.6--0.8   & 4--14\%\\
\end{tabular}
\end{ruledtabular}
\end{table}
\section{Encoded injection from Bacon--Shor to surface code}
\label{sec:switch}
The transmission code need not be the computation code, provided that the
logical Pauli algebra can be transferred across the boundary. We implement the
BS--SC boundary as an encoded-state injection. Its geometry is that of magic-state
injection on the rotated surface code~\cite{lao2022}, a small encoded patch
embedded in a larger one and grown to full distance, but with a crucial
difference. Magic-state injection prepares a known resource state whose
one-shot, non-fault-tolerant injection is tolerable because the state is later
distilled, whereas the repeater injects an \emph{arbitrary, unknown} logical
state that cannot be re-prepared or distilled. The injection therefore has to
preserve the logical information fault-tolerantly, which we realize as a
fault-tolerant code deformation (gauge fixing) that activates the surface-code
stabilizers over $\Theta(d)$ rounds while the Bacon--Shor gauges remain measured,
rather than as a single-shot activation. The
incoming $\mathrm{BS}(m\times n)$ block is embedded into the
first $n$ columns of an $m\times m$ rotated surface-code patch, and the remaining
$m-n$ columns are filled with $\lvert+\rangle$ resource qubits. We choose this
orientation so that the high-loss $Z$-sector logical, of Bacon--Shor distance
$d_Z=m$, is already a full length-$m$ surface-code $Z$ string, while the
lower-loss $X$-sector logical (length $n$) is the one extended across the added
$\lvert+\rangle$ columns. This is why the resource is prepared in the $X$ basis.
The surface-code checks are then activated, projecting the enlarged system into a
definite surface-code syndrome sector.

Algebraically, let $G_{\rm BS}$ be the Bacon--Shor gauge group, $S_{\rm res}$ the
stabilizer group of the added resource qubits, and $S_{\rm meas}$ the stabilizer
relations generated by the activated surface-code measurements. The interface is
valid if the Bacon--Shor logical Pauli pair is mapped injectively to the
surface-code logical Pauli pair, namely
\begin{equation}
 \bar X_{\rm BS}\bar X_{\rm SC},\;\bar Z_{\rm BS}\bar Z_{\rm SC}
 \in \langle G_{\rm BS},S_{\rm res},S_{\rm meas}\rangle .
 \label{eq:interface-rowspace}
\end{equation}
Here the bracket denotes equivalence in the binary Pauli module. Representatives
may be multiplied by Bacon--Shor gauge generators, resource stabilizers, and
measured surface-code stabilizer relations. The Bacon--Shor gauge degrees of
freedom are not part of the protected logical subsystem.
In the embedding used here the length-$m$ column logical $\bar Z_{\rm BS}$ already
coincides with the surface-code $Z$ logical, $\bar Z_{\rm BS}=\bar Z_{\rm SC}$,
while the length-$n$ row logical $\bar X_{\rm BS}$ is extended to the surface-code
$X$ logical through the $m-n$ resource columns, the resource being prepared so
that $\bar X_{\rm res}=\bar X_{\rm BS}\bar X_{\rm SC}\in S_{\rm res}$. Binary
stabilizer-tableau reduction verifies Eq.~\eqref{eq:interface-rowspace} for
$\mathrm{BS}(m\times3)\to\mathrm{SC}(m\times m)$ with $m=5,7,9$: the
post-measurement centralizer quotient contains exactly one anti-commuting logical
pair, and the boundary outcomes fix only gauge or Pauli-frame data rather than
measuring the input logical qubit (Appendix~\ref{app:switch-verification}).

The reverse $\mathrm{SC}(m\times m)$--to--$\mathrm{BS}(m'\times m)$ switch used before the return transmission is the same multi-round deformation applied to the other sector, extending the transmitted $Z$-distance from $m$ to $m'$, and obeys the same logical-algebra criterion. Both interfaces therefore grow the code. A size-reducing contraction, which would relinquish distance rather than the encoded qubit, does not arise in a round trip built only to add protection. In the
present simulations both interfaces are modeled as ideal logical maps once the
row-space and centralizer tests are satisfied.

Fault tolerance follows by co-measuring same-type Bacon--Shor gauges and
surface-code stabilizers during a $\Theta(d)$ activation window before releasing
the gauges~\cite{horsman2012,vuillot2019}. This closes the detector gap of an
abrupt switch while transferring protection continuously between the two codes.
The endpoint erasure distances are $(m,n)$ and $(m,m)$ for the tested
$m=5,7,9$, and Stim's shortest-graphlike-error diagnostic gives stitching-window
distance $d=m$ in both sectors for $m=3,5,7$, whereas an abrupt switch gives
$d=1$. The schedule and a worked $\mathrm{BS}(5\times3)\to\mathrm{SC}(5\times5)$
example are given in Appendix~\ref{app:switch-verification}. This construction realizes a fault-tolerant interface between a subsystem Bacon--Shor code and a topological surface code for an arbitrary unknown logical qubit. In the round-trip simulations the boundary is consequently treated as a logically transparent
interface; embedding both multi-round switches into the full circuit-level round
trip remains future work.

\section{Deferred-decoding distributed computation}
\label{sec:deferred}
\begin{figure*}[t]
\centering
\includegraphics[width=0.70\linewidth]{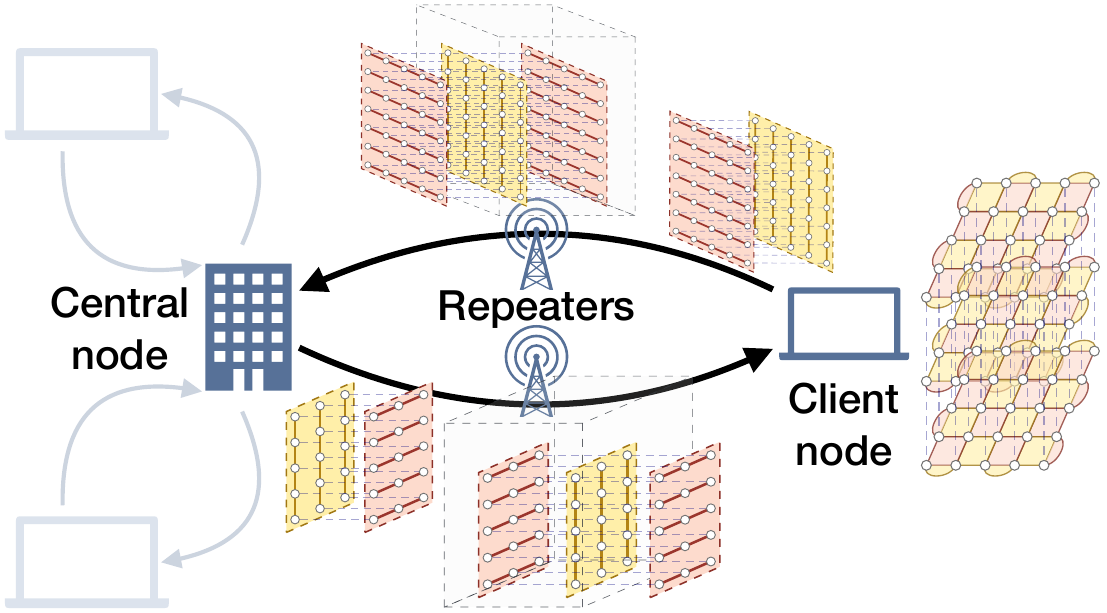}
\caption{The BS--SC--BS deferred-decoding round trip. The center prepares the
logical qubit in a transmission-optimized asymmetric Bacon--Shor code
$\mathrm{BS}_1(m\times n)$ ($d_Z=m\gg d_X=n$) and transmits it to the client. The
channel loss enters as the transmitted-sector erasure rate $p_Z$
[Eq.~\eqref{eq:rates}]. The client injects the incoming state into a surface code
$\mathrm{SC}(m\times m)$ through the fault-tolerant switch of the previous section,
runs the computation (Pauli rate $p_{\rm comp}$), switches to a return code
$\mathrm{BS}_3(m'\times m)$, and transmits it back ($p_Z$ again). The client never
runs the global decoder. It forwards its measurement (syndrome) records to the
center, which assembles the full heterogeneous record into one combined complex
and decodes it in a single deferred joint pass. The two Bacon--Shor legs carry the
large transmitted-sector distance ($m$ and $m'$). The surface code serves only as
the computation block.}
\label{fig:bsspipe}
\end{figure*}

The same construction extends from a one-way link to a round trip
(Fig.~\ref{fig:bsspipe}), and this is
where the foliated, code-switching view pays off architecturally. Consider a
central node that delegates a computation to a remote client node. The
central node prepares the logical qubit in a transmission-optimized asymmetric
Bacon--Shor code and sends it to the client (phase~1). The client injects the
incoming state into a surface code using the BS--SC interface above, runs the
computation (phase~2), switches back to a Bacon--Shor transmission code, and
returns the qubit to the center (phase~3), where all decoding is performed. We
perform the deferred joint decode. The central node assembles the full
heterogeneous detector record (outbound Bacon--Shor, surface-code computation, and
return Bacon--Shor, stitched through the two interfaces) into a single combined
complex and decodes it in one pass with the maximum-likelihood erasure decoder. We find this joint decode equals the phase-separated result exactly, in
both sectors and at all losses (e.g.\ round-trip $P_{\rm succ}\simeq0.99$ at
$p_{\rm ch}=0.10$, $\eta_r=0.9$, dominated by the local $X$ sector of distance
$d_X=n$). This equality is expected and is the correct physics. The transmission
legs are causally sequential teleportation steps, so an erasure that defeats one
leg propagates through the re-encoded logical and cannot be repaired by a later
leg's syndrome. There is no cross-leg gain to be had, and the joint decode neither
beats nor underperforms decoding the legs in sequence. The significance of
deferred decoding is therefore architectural rather than a decode-performance gain.
The client need not run the global QEC decoder. It executes the
prescribed preparations, measurements, and logical operations, records the
outcomes, and forwards the syndrome record to the central node. All global error
correction is deferred to the center, a substantial simplification for a remote
or lightweight station. Where the decode runs is itself a design choice that the locality of
distributed computation opens up. For a small enough computation block a client could instead
fold the loss and errors of its incoming leg into a light local decode of its own,
trading the decoder-free client for a lighter, local one. For the Clifford pipeline considered here, deferring the global decode incurs no decoding-performance penalty. The
operations are Clifford, so an error incurred in any phase propagates
deterministically and is recorded by the downstream detectors, and the central
decoder inverts this propagation from the complete record. The only events it
cannot repair are logical errors. But a logical error that would defeat
the round trip is equally beyond repair by an intermediate correction, since it
commutes with every stabilizer and leaves no syndrome to act on. Decoding in one
deferred pass is therefore no weaker than correcting after each leg. It simply
relocates the global decoder to the center and frees the client of it.

\begin{figure*}[t]
\centering
\includegraphics[width=0.92\linewidth]{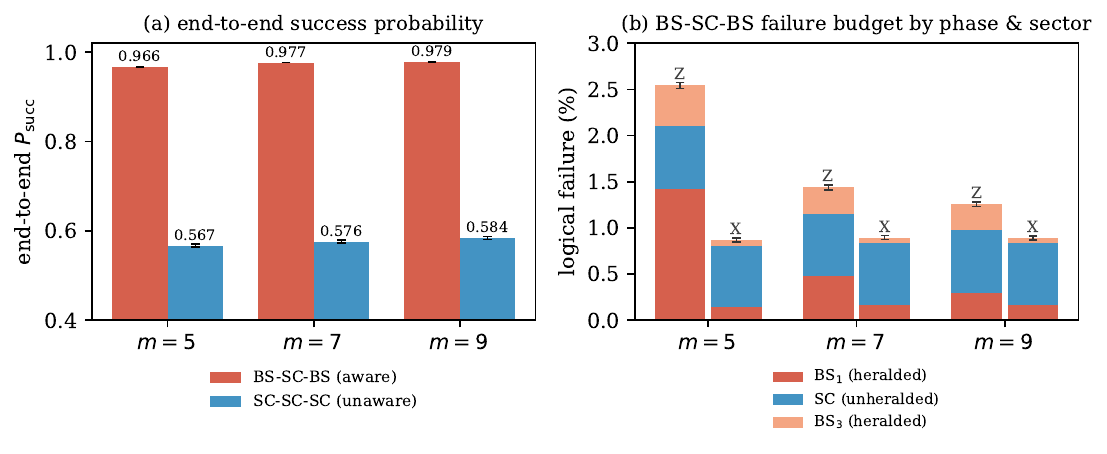}
\vspace{-8pt}
\caption{Deferred-decoding round trip at $p_{\rm ch}=0.10$, $p_{\rm comp}=0.01$, $\eta_r=0.9$, one repeater per leg ($\mathrm{BS}(m\times3)\to\mathrm{SC}(m\times m)\to\mathrm{BS}(m'\times m)$,
$m'=7,9,11$ for $m=5,7,9$).
(a)~Lumped end-to-end $P_{\rm succ}$, BS--SC--BS versus the transmission-unaware baseline (SC--SC--SC).
(b)~The same BS--SC--BS failure resolved into its constituents. For each $m$ the two
bars give the logical-$Z$ and logical-$X$ failure, each decomposed by phase. The
two Bacon--Shor transmission legs (BS$_1$, BS$_3$) fail by \emph{heralded}
erasure (the decoder knows it failed, so the event is post-selectable), whereas
the surface-code computation (SC) fails by an \emph{unheralded} logical Pauli
error that silently corrupts the result. A single scalar $P_{\rm succ}$ lumps these two
operationally distinct mechanisms. The decomposition shows that the heralded
Bacon--Shor contribution is suppressed by scaling $m$, while the unheralded
surface-code contribution is flat and comes to dominate the residual in both
sectors. Statistics: $2\times10^5$ Monte Carlo samples for BS--SC--BS and
$2\times10^4$ for SC--SC--SC. Error bars are $\pm1$ standard error and are smaller
than the markers in (a) and comparable to the line widths in (b).}
\label{fig:bss-breakdown}
\end{figure*}

This exact deferred-decoding equivalence applies to the Clifford pipeline
considered here. Universal computation additionally requires a gadget-local,
distance-$d$ feedforward decode at each non-Clifford gate to determine its
conditional Clifford correction. Thus the client is relieved of the global QEC
decoder, not of all local processing. Whether this local feedforward can be
separated completely from the upstream deferred record remains an open
circuit-level question; the Pauli-frame argument is detailed in
Appendix~\ref{app:nonclifford}.

Concretely the pipeline is
\[
 \mathrm{BS}(m\times n)\;\to\;\mathrm{SC}(m\times m)\;\to\;\mathrm{BS}(m'\times m),
\]
with $n<m<m'$. The outbound and return legs use loss-matched asymmetric
Bacon--Shor codes (Sec.~\ref{sec:perhop}), while the computation runs on a symmetric surface
code. The return dimension $m'$ is chosen by the same code-selection principle so
that the return-leg failure is comparable to the outbound and computation
contributions. In practice it may be larger because the returned logical frame is
already correlated with the syndrome record accumulated in the first two phases.
We call this the transmission-aware pipeline, BS--SC--BS, since its legs
are matched to the channel asymmetry. The natural alternative, a
\emph{transmission-unaware} baseline that uses the surface code throughout
(SC--SC--SC), is what one would build without code switching.

The simulation reported here is deliberately the simplest feasibility setting.
Each transmission leg carries a single intermediate repeater (a five-layer $ZXZXZ$ foliation), and the computation block is a single surface-code layer
subject to depolarizing noise at rate $p_{\rm comp}$. We do not simulate explicit
logical gates. The computation phase is modeled as a noisy memory channel whose
error budget is set by $p_{\rm comp}$. This is sufficient to test what the round
trip requires, that the heterogeneous BS$\to$SC$\to$BS record decode jointly and
survive a representative computation-phase error. This suffices as a feasibility check of the heterogeneous round-trip decode.

Figure~\ref{fig:bss-breakdown}(a) reports the round-trip effective transmission
rate. At $p_{\rm ch}=0.10$, $p_{\rm comp}=0.01$ ($\eta_r=0.9$, one repeater per leg), BS--SC--BS reaches $P_{\rm succ}\simeq0.98$ at $m=9$, versus $0.58$ for
SC--SC--SC, a $\sim\!30\times$ reduction in logical-$Z$ failure (and
$\sim\!15\times$ already at $m=5$). This holds at equal data-qubit count. With
$N_{\rm aware}=mn+m^2+m'm$ and $N_{\rm unaware}=3m^2$ ($m'=7,9,11$ for $m=5,7,9$),
BS--SC--BS uses no more data qubits than SC--SC--SC, so the advantage is not due to
a larger code. The
all-surface-code pipeline is held back for the reason identified in Sec.~\ref{sec:perhop}.
Its two transmission legs use the surface code, a weaker erasure-transmission
block than the gauge-rich Bacon--Shor code, and the round trip compounds this
transmission bottleneck with the computation phase. Replacing the two
transmission legs with loss-matched Bacon--Shor codes removes this bottleneck,
leaving the surface code to play its natural role as the computation block.

The finite-size scaling is consistent with sub-threshold operation. With the
computation idealized ($p_{\rm comp}=0$), the round-trip logical-$Z$ failure falls
with code size (at $p_{\rm ch}=0.10$ it is $1.88\%,\,0.71\%,\,0.47\%$ for
$m=5,7,9$, and the suppression strengthens at higher channel loss), so errors are
removed by scaling rather than accumulated by the round trip.

Resolving the failure by phase isolates where the budget is spent
[Fig.~\ref{fig:bss-breakdown}(b)]. For $m=9$ at
$p_{\rm ch}=0.15$, $p_{\rm comp}=0.01$, the logical-$Z$ failure splits into
$0.41\%$ (outbound transmission), $0.78\%$ (surface-code computation), and
$0.38\%$ (return transmission), totaling $1.56\%$ (with $0.86\%$ for the
$X$ sector). Once the elongated Bacon--Shor blocks suppress the $Z$-sector
transmission bottleneck, no single phase dominates. The surface-code computation
is now the largest single contributor, comparable to the two transmission legs
combined, and the lower-distance $X$ sector remains well controlled because its
syndromes are measured locally. In particular the return leg does not dominate,
so the round trip is not bottlenecked by the cost of sending the qubit back for
central decoding.

The decomposition also separates two operationally distinct failure mechanisms
that the single $P_{\rm succ}$ figure of merit lumps together. The Bacon--Shor transmission
legs fail by \emph{heralded} erasure. The row-space decoder reports when the
erased pattern is undecodable, so such an event is flagged and can be
post-selected or retransmitted, costing throughput rather than fidelity. The
surface-code computation, in contrast, fails by an \emph{unheralded} logical
Pauli error that commutes with every stabilizer and silently corrupts the
output. Scaling $m$ suppresses the heralded Bacon--Shor contributions but leaves
the unheralded surface-code term essentially flat, so for large $m$ the residual
round-trip failure is dominated by the computation phase, the contribution that
genuinely limits the fidelity of the delivered logical qubit. Reporting $P_{\rm succ}$
alongside this decomposition therefore states both the aggregate success
probability and the operationally relevant residual.

The decisive point is that this round-trip overhead is modest. Deferring the
global decoding to the center costs one return transmission yet leaves
$P_{\rm succ}\simeq0.98$, more than an order of magnitude lower in logical
failure than the all-surface-code alternative. With the deferred joint decode of
the full heterogeneous record carried out as above, and equal to the
phase-separated result for these sequential legs, distributed quantum computation
with a \emph{decoder-free client} for the global QEC task (computation at the edge,
error correction at the center) is viable within the same foliated-repeater
architecture, reusing the asymmetric Bacon--Shor transmission code of Sec.~\ref{sec:perhop} for both legs of the trip. One remaining idealization is the noiseless interface map. The fault-tolerant injection schedule of the preceding section
shows this is well-founded, and a fully circuit-level treatment of the two
switches embedded in the round trip is the natural next step.

\section{Discussion}
Viewing the measurement-based repeater as a foliated code turns the choice of
encoding into an asymmetry-matching problem. Because the foliation axis is the
transmission direction, the dominant channel loss is concentrated on one sector, and a
code with a correspondingly lopsided distance spends its protection where it is
needed.

Underlying this is a structural fact about DQC. A distributed computation can
teleport its logical data over classical channels once entanglement is in place,
but the entanglement that teleportation consumes can only be distributed,
never synthesized by local operations and classical communication. Loss-tolerant
quantum transmission is therefore an irreducible component of any distributed
computation rather than a one-off preprocessing step. The loss asymmetry we exploit
is, moreover, precisely the split between this transmitted, entanglement-carrying
sector and the locally prepared sector that classical instructions could in
principle replace. So treating transmission and computation within a single
measurement-based record puts the two on the footing this structure implies.

We stress what the claim is and is not. We do not advocate a fixed pairing of a
Bacon--Shor transmission code with a surface-code computation. The message is
structural. Against the backdrop that DQC needs high-fidelity entanglement
distribution, the task a quantum repeater is built for, and that recently proposed
measurement-based repeaters carry an intrinsic loss asymmetry, our point is
that one should \emph{transmit with a code built for transmission}: a code whose
distance asymmetry is matched to the channel yields better transmission than a
symmetric code of equal size. The framework then has three movable parts: (i)~at the
nodes, a fault-tolerant code switch lets each stage run in the code best suited to
it---a computation-friendly code while computing, a transmission-friendly code for
each (re)transmission leg; (ii)~deferring the global decode to a single node then
removes the decoder requirement at the client; or (iii)~exploiting the locality of
DQC, for small enough computation blocks each client may instead perform its own lighter decode, clearing the errors of its incoming leg and its local computation before they propagate into the return transmission, rather than returning the record to a central node. Which
specific codes and which placement of the decoding are optimal is left open. The
framework, not a particular code pair, is the contribution.
\section{Outlook}
Several idealizations bound these conclusions and mark the next steps. The
computation phase is modeled as a single noisy surface-code layer subject to
depolarizing noise, a memory channel rather than an explicit logical-gate
circuit, so the round-trip simulation is a feasibility check at the simplest
operating point (one intermediate repeater per leg). The Bacon--Shor/surface-code interface map is treated as
noiseless in the transport simulation. The fault-tolerant injection schedule
shows this is well-founded, but folding both switches into a single
circuit-level Monte Carlo is the natural way to remove the remaining
idealization. The transmission analysis, like the loss model it
adopts~\cite{niu2023}, treats photon loss as heralded erasure, the dominant
photonic error, and does not include unheralded Pauli errors on the transmitted
and local layers. Including them would reinstate a finite threshold set by the
code distances $(d_Z,d_X)=(m,n)$, so the threshold-free property is specific to the loss channel.

A broader opportunity is to generalize beyond the measurement-based repeater of Ref.~\cite{niu2023}. Any all-photonic repeater~\cite{azuma2015,pant2017,borregaard2020,azuma2023} distributes entanglement by preparing, propagating, and measuring an encoded photonic state, itself a measurement-based error-correction process, so the deferred-decoding round trip developed here should extend to all-photonic repeaters at large. Realizing this would turn the idealizations above into substantive programs, in particular fault-tolerant interfaces between arbitrary code pairs at each junction and a circuit-level treatment of non-Clifford operations within the deferred-decoding picture.

Beyond this, the asymmetry-matching principle invites other
biased-distance codes and biased-noise hardware (tuning surface-code aspect ratios~\cite{tuckett2018}, the XZZX code~\cite{bonilla2021}, and biased fusion networks~\cite{sahay2023}), and integration with a
protocol-level cost model would let the code be co-optimized with repeater
spacing across an end-to-end network. A further opportunity lies below this
data-qubit accounting. Because the Bacon--Shor gauge operators are weight two,
each foliated layer is a low-degree graph (Fig.~\ref{fig:foliation}), so its
photonic resource cluster needs fewer entangling fusions per qubit than the
higher-weight surface code, a plausible implementation advantage in the spirit of fusion-based architectures~\cite{bartolucci2023} (simpler
preparation, lower preparation-stage loss) that our loss model, which counts only
data qubits and per-hop loss, does not capture and that merits a dedicated
resource-state analysis.
\begin{acknowledgments}
This research was supported by Korea Institute of Science and Technology Information (KISTI) (K26L1M3C5). This research was supported by the National Research Council of Science \& Technology (NST) grant funded by the Korean government (MSIT) (No.~CAP22055-000).
\end{acknowledgments}

\appendix
\section{Gauge-structure origin of the Bacon--Shor erasure advantage}
\label{app:gauge}
\begin{figure*}[t]
\centering
\includegraphics[width=0.92\linewidth]{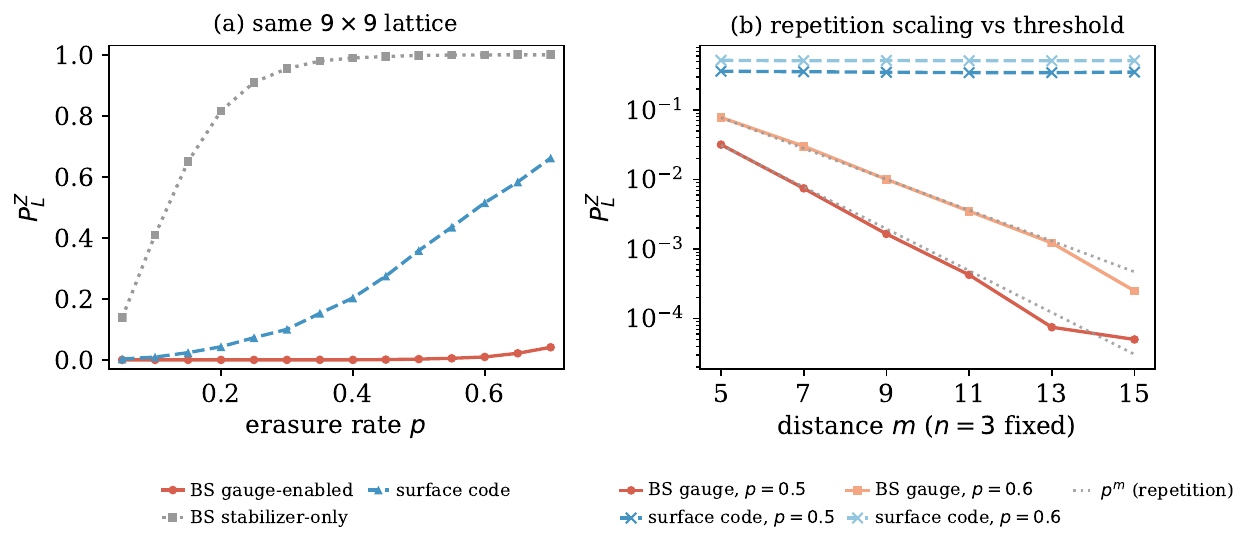}
\vspace{-6pt}
\caption{Gauge-structure origin of the Bacon--Shor erasure advantage (i.i.d.\
data erasure, perfect syndrome, $Z$-sector). The control parameter $p$ is a
uniform erasure probability applied independently to each data qubit (syndrome
qubits assumed intact), isolating the code structure. In the repeater the
transmitted $Z$-sector operates near $p=p_Z$ [Eq.~\eqref{eq:rates}].
(a)~On the same $9\times9$
lattice, removing the gauge generators from the row-space equivalence
(``stabilizer-only,'' a mechanism diagnostic rather than a code baseline)
collapses the Bacon--Shor performance from $P_L^Z\approx0$ to $P_L^Z\approx1$,
below the surface code. The full gauge-enabled decoder is near-deterministic up
to $p\simeq0.5$. (b)~$P_L^Z$ versus distance $m$ ($n=3$ fixed) at $p=0.5$ and $0.6$, erasure
rates lying at and above the surface code's $\sim\!50\%$ percolation
threshold. At both rates the gauge-enabled Bacon--Shor $Z$-sector follows the
repetition-code scaling $p^{\,m}$ (solid lines, hugging the $p^m$ guides),
whereas the surface code ($\times$) is saturated and flat in distance. The
Bacon--Shor sector keeps suppressing the failure exponentially exactly where the
surface code cannot. Only the dominant transmitted
$Z$-sector is shown. The local $X$-sector obeys the same mechanism under
$m\leftrightarrow n$ and stays benign at the small local erasure rate $p_X$.}
\label{fig:gauge}
\end{figure*}
As an independent decoder validation, direct i.i.d.\ erasure on
$\mathrm{SC}(9)$ gives
$P_L^Z=2\times10^{-4},\,0.10,\,0.35$ at
$p=0.01,\,0.30,\,0.50$, respectively, reproducing the expected
$\sim\!50\%$ surface-code erasure threshold. Clean erasure-only simulations of
the isolated Bacon--Shor transmission block likewise confirm that the exponent
is controlled by $d_Z=m$: at fixed $m$, the $Z$-sector failure is nearly
independent of the transverse dimension $n$ and falls as $\sim p_Z^{\,m}$.

Two distinct effects combine in the equal-size comparison of Fig.~\ref{fig:n81} of the main text. The gap
$\mathrm{SC}(9\times9)\!\to\!\mathrm{BS}(9\times9)$ is a code-family
effect. The further gain $\mathrm{BS}(9\times9)\!\to\!\mathrm{BS}(27\times3)$ is
the aspect-ratio effect. The former is by far the larger (at the heaviest
loss it lifts $P_{\rm succ}$ from $0.23$ to $0.89$, against a further $0.89\to0.99$
from the aspect ratio). A diagnostic row-space test isolates the origin
of the first. Decoding the same Bacon--Shor lattice with the gauge generators removed from the equivalence, so that the logical is deformable only by the high-weight stabilizers as in a subspace code, collapses the $Z$-sector
from $P_L^Z\!\approx\!0$ to $P_L^Z\!\approx\!1$, worse than even the surface code
[Fig.~\ref{fig:gauge}(a)]. The Bacon--Shor advantage is therefore a
gauge-structure effect. The gauge freedom supplies many equivalent dressed
representatives of each logical, so that the leading erasure failure follows the
repetition-code scaling $P_L^Z\sim p_Z^{\,m}$ [Eq.~\eqref{eq:detour}], confirmed
directly over $m=5$--$15$ at the two erasure rates $p=0.5$ and $0.6$
[Fig.~\ref{fig:gauge}(b)], that is, at and above the surface code's
$\sim\!50\%$ erasure threshold. The surface-code logical, by contrast, is
deformable only by its stabilizers and is limited by that percolation-type
threshold. At the same two erasure rates its $P_L^Z$ is saturated and
essentially flat in distance, so enlarging the code cannot reduce it. The
Bacon--Shor sector thus keeps suppressing the failure exponentially in $m$
precisely in the high-loss regime where the surface code can no longer, the
operating regime of a sparsely repeatered link. The
aspect-ratio effect then operates within the Bacon--Shor family, matching
$d_Z\!=\!m$ to the high-loss transmitted sector. We display the dominant transmitted $Z$-sector throughout. The local $X$-sector is governed by the same gauge mechanism under $m\leftrightarrow n$ and, at the small local erasure rate $p_X$, contributes negligibly to $P_L$.

\section{Long-distance performance at other repeater efficiencies}
\label{app:longdist}
\begin{figure*}[t]
\centering
\includegraphics[width=0.92\linewidth]{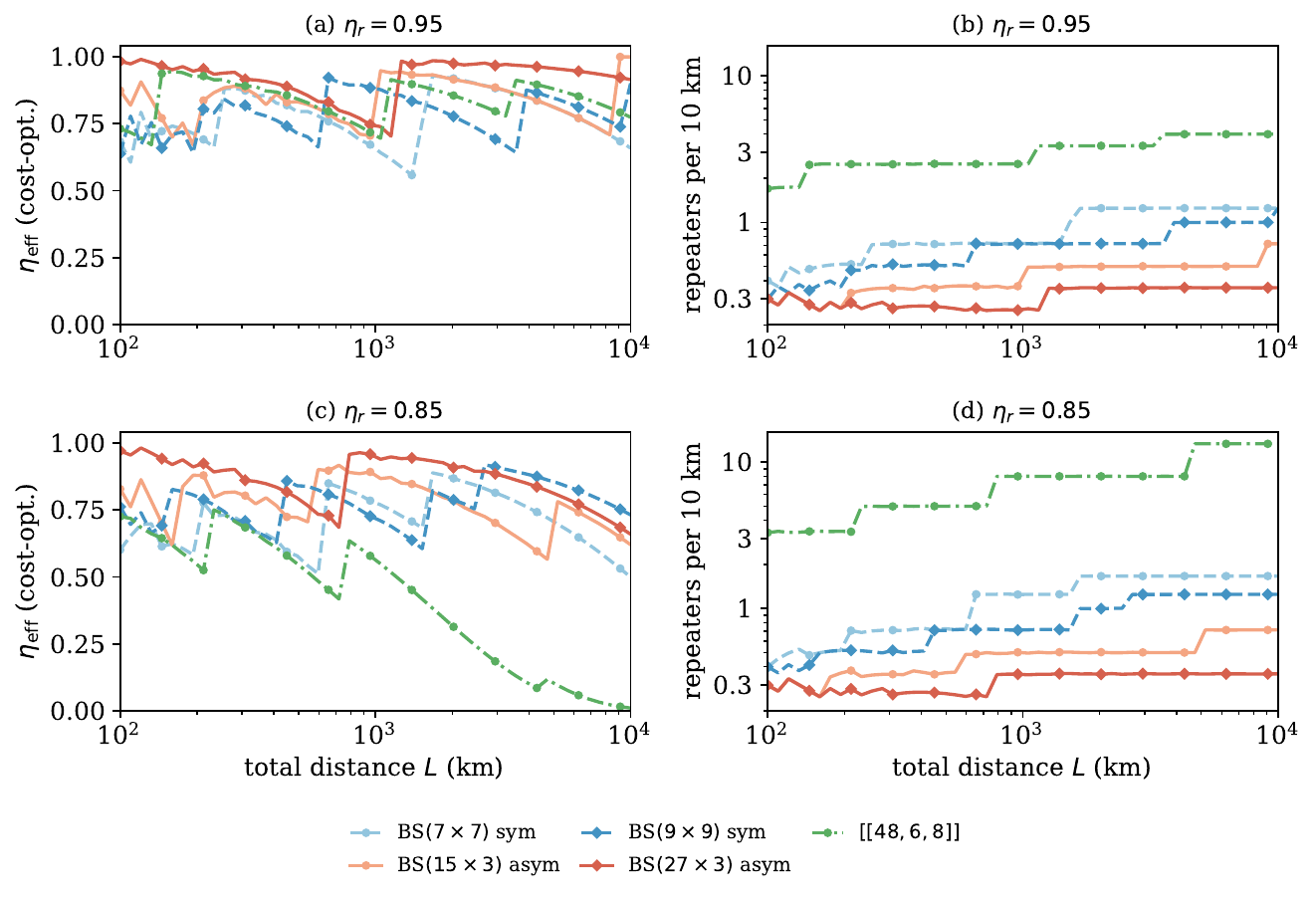}
\vspace{-7pt}
\caption{Long-distance performance at the other two repeater efficiencies of
Ref.~\cite{niu2023}, complementing the $\eta_r=0.9$ case of
Fig.~\ref{fig:longdist}. Top row (a),(b) is $\eta_r=0.95$ (local loss
$p_X=1-\sqrt{\eta_r}\approx0.025$). Bottom row (c),(d) is $\eta_r=0.85$
($p_X\approx0.078$). Panels (a),(c) give the cost-optimal effective transmission
rate $\eta_{\rm eff}$ versus total distance $L$. Panels (b),(d) give the corresponding cost-optimal
repeater density (repeaters per $10$~km). Four equal-pair codes and the $[[48,6,8]]$ benchmark of Ref.~\cite{niu2023} (green, dash-dotted), all under the same maximum-likelihood erasure decoder, are compared. The pairs are the
$\sim\!48$-qubit pair $\mathrm{BS}(7\times7)$ (symmetric) and
$\mathrm{BS}(15\times3)$ (asymmetric), and the $81$-qubit pair
$\mathrm{BS}(9\times9)$ (symmetric) and $\mathrm{BS}(27\times3)$ (asymmetric). At
matched code size the asymmetric codes (warm colors, solid) sit at the higher $\eta_{\rm eff}$ in (a),(c) and the lower repeater density in (b),(d), roughly
half the stations of their symmetric counterparts (cool colors, dashed), across the whole range. Lowering $\eta_r$ from $0.95$ to $0.85$ raises the local loss $p_X$ and
shifts every curve toward denser spacing, but leaves the asymmetric advantage and its ordering intact. The benchmark remains pinned to an order-of-magnitude denser spacing at both efficiencies (logarithmic density axis), its $\eta_{\rm eff}$ additionally collapsing at long distance for $\eta_r=0.85$. Markers pair resource-matched codes, circles for the $\sim\!48$-qubit codes and diamonds for the $81$-qubit pair. The discontinuities are the integer steps of the discrete
repeater-number optimization, as in Ref.~\cite{niu2023}. Per-hop rates are
maximum-likelihood erasure Monte Carlo in the reliable-syndrome limit,
and the cascade follows Eq.~\eqref{eq:cascade}.}
\label{fig:longdist-app}
\end{figure*}
Figure~\ref{fig:longdist} of the main text fixes the repeater efficiency at
$\eta_r=0.9$. Figure~\ref{fig:longdist-app} repeats the
analysis at $\eta_r=0.95$ (top row) and $\eta_r=0.85$ (bottom row), the other two
values reported in Ref.~\cite{niu2023}. The qualitative ordering is unchanged at every efficiency.
The asymmetric Bacon--Shor codes are cost-optimal at a sparser repeater spacing
than their equal-size symmetric counterparts, and the higher local-loss case
($\eta_r=0.85$) simply shifts every curve toward denser spacing without altering
the asymmetric advantage.

\section{Robustness to the syndrome loss model}
\label{app:syndrome}
The main text follows Ref.~\cite{niu2023} in treating all syndrome (ancilla)
qubits as locally generated and measured, so that they experience only the local
erasure rate $p_X=1-\sqrt{\eta_r}$. This is an assumption of the benchmark model,
adopted for a matched comparison rather than a physical necessity, and in some
hardware the transmitted-layer syndromes must themselves cross the link and would
then incur the transmitted rate. The comparison proves robust to this assumption
from both sides, as shown below.

\subsection{Cross-layer syndrome redundancy and the reliable-syndrome limit}
\label{app:crosslayer}
The long-distance results of the main text use an effective foliated model in
which the transmitted-sector syndrome is taken reliable. Here we justify that
limit and exhibit the cross-layer depth it requires, using the validated
two-dimensional erasure decoder. In the foliated cluster each transmitted-sector
stabilizer is re-measured in every layer it threads. Modeling such a check as
available unless its locally generated ancilla is erased in all $T$ of
those layers gives an availability $1-p_X^T$. Table~\ref{tab:crosslayer} reports
the per-hop $Z$-sector erasure failure $P_L^Z$ as a function of $T$ at a
representative high per-segment loss ($p_Z=0.64$, $\eta_r=0.9$, $p_X\approx0.05$),
with the transmitted data erased once.

A single layer ($T=1$, one-shot syndrome) over-penalizes the elongated codes.
$\mathrm{BS}(27\times3)$ is then no better than the symmetric
$\mathrm{BS}(9\times9)$, which is the origin of the apparent syndrome-loss
fragility noted in the main text. Already at $T=2$--$3$, however, $P_L^Z$
converges to its reliable-syndrome ($T\to\infty$) value and the asymmetric
advantage is fully restored ($P_L^Z\approx3\times10^{-4}$ for
$\mathrm{BS}(27\times3)$ against $1.7\times10^{-2}$ for $\mathrm{BS}(9\times9)$).
The reliable-syndrome limit used in Fig.~\ref{fig:longdist} and
Table~\ref{tab:niu-compare} is therefore reached at the modest cross-layer depth
that only a few foliation layers already supply. As an independent check, a
foliated decode built from the Bacon--Shor reduction to two repetition
codes~\cite{pesah2025} and run as a Stim memory with explicit local-ancilla loss
(rather than reliable syndrome) likewise preserves the asymmetric advantage at
high loss. At $p_Z\approx0.78$ the asymmetric $\mathrm{BS}(27\times3)$ retains
$P_{\rm succ}\approx0.99$ while the symmetric $\mathrm{BS}(9\times9)$ falls to
$\approx0.67$.

\begin{table}[t]
\caption{Per-hop transmitted-sector erasure failure $P_L^Z$ versus cross-layer
depth $T$ (each check available with probability $1-p_X^T$), at $p_Z=0.64$,
$\eta_r=0.9$ ($p_X\approx0.05$). $2\times10^4$ samples, validated row-space
decoder. $T=1$ is the single-layer (one-shot) syndrome. $T\to\infty$ is the
reliable-syndrome limit used in the main text.}
\label{tab:crosslayer}
\begin{ruledtabular}
\begin{tabular}{lcccc}
code & $T=1$ & $T=2$ & $T=3$ & $T\to\infty$\\
\hline
$\mathrm{BS}(9\times9)$ (sym.)   & 0.121 & 0.025 & 0.017  & 0.017\\
$\mathrm{BS}(15\times3)$ (asym.) & 0.124 & 0.007 & 0.002  & 0.001\\
$\mathrm{BS}(27\times3)$ (asym.) & 0.150 & 0.005 & 0.0003 & 0.000\\
\end{tabular}
\end{ruledtabular}
\end{table}

\subsection{Conservative channel-exposed syndromes}
\label{app:asis}
A more conservative, and in some hardware more
realistic, variant assigns the transmitted layers' syndromes the same erasure
rate $p_Z$ as the data they read out. This penalizes the lower-distance sector,
since its syndrome is then degraded by the link, and lowers every code's logical success probability. It does not, however, alter any qualitative conclusion. This
analysis should therefore be read as a stress test of the architectural claim
rather than as the baseline Niu loss model.
Table~\ref{tab:asis} repeats the equal-size comparison of Fig.~\ref{fig:n81}
under this conservative model (with the same corrected foliated primal/dual
erasure treatment for all codes). The surface code remains a poor transmission
block, the symmetric Bacon--Shor code degrades at high loss, and the asymmetric
$\mathrm{BS}(27\times3)$ again sustains the highest $P_{\rm succ}$. This supports the
robustness of the main-text conclusion that the asymmetric-code advantage is not
an artifact of the local-syndrome convention.

\begin{table}[t]
\caption{Logical success probability under the conservative loss model in which
transmitted-layer syndromes also incur the transmitted erasure rate $p_Z$
($N=81$, $\eta_r=0.9$, one repeater). Compare with Fig.~\ref{fig:n81} (main-text,
local-syndrome model).}
\label{tab:asis}
\begin{ruledtabular}
\begin{tabular}{lcccc}
 & $p_{\rm ch}{=}0.10$ & $0.20$ & $0.30$ & $0.45$\\
\hline
$\mathrm{SC}(9\times9)$ (sym.)   & 0.763 & 0.564 & 0.395 & 0.205\\
$\mathrm{BS}(9\times9)$ (sym.)   & 0.987 & 0.969 & 0.933 & 0.802\\
$\mathrm{BS}(27\times3)$ (asym.) & 0.987 & 0.977 & 0.962 & \textbf{0.937}\\
\end{tabular}
\end{ruledtabular}
\end{table}

\section{Stabilizer and spacetime-distance verification of the BS--SC switch}
\label{app:switch-verification}

The logical-algebra condition of Eq.~\eqref{eq:interface-rowspace} was verified
by binary stabilizer-tableau row reduction. For
$\mathrm{BS}(m\times3)\to\mathrm{SC}(m\times m)$ with $m=5,7,9$, the column
$\bar Z$ coincides in the two codes, $\bar X_{\rm BS}\bar X_{\rm SC}$ is
supported only on the resource columns, and the post-measurement centralizer
quotient contains exactly one anti-commuting logical pair. The bare
$\bar X_{\rm BS}$ anticommutes with a boundary $Z$-stabilizer, which is precisely
why its representative is extended rather than measured. First-round boundary
outcomes may be random, as in surface-code state injection, but they enter only
as gauge or Pauli-frame data: the measured boundary check promotes the truncated
Bacon--Shor representative to the full surface-code representative without
revealing either member of the input logical pair.

The pre-switch configuration, the Bacon--Shor block with product-state resource
columns, has erasure distances $(d_Z,d_X)=(m,n)$, while the post-switch surface
code has $(m,m)$. Thus the minimum dressed-logical weight equals or exceeds the
Bacon--Shor design value $\min(m,n)=n$ at both endpoints, as checked for
$m=5,7,9$. The transient distance is instead the spacetime fault distance of the
measurement schedule. Within a CSS sector, same-type Bacon--Shor gauges and
surface-code stabilizers commute: vertical $ZZ$ gauges commute with
$Z$-plaquettes and horizontal $XX$ gauges with $X$-plaquettes, whereas checks of
opposite type can anticommute. We therefore co-measure the same-type gauges and
newly activated stabilizers for a short window of rounds, and release the gauges
only after the surface-code syndrome history has been established. This pins each new stabilizer to the incoming gauge record and closes the detector that an abrupt switch would leave open. The activation is a gauge-fixing step in the sense of Ref.~\cite{vuillot2019}, Bacon--Shor and surface codes being two limits of the compass-code family~\cite{li2019compass}.

For a concrete example, consider the $Z$ sector of
$\mathrm{BS}(5\times3)\to\mathrm{SC}(5\times5)$. The $15$ data qubits occupy
columns $0$--$2$, ten resource qubits fill columns $3$--$4$, and the logical
$\bar Z$ is the shared weight-$5$ string on column $0$. The Bacon--Shor block has
12 vertical $ZZ$ gauges and the surface code adds 12 $Z$-plaquettes. During the
activation round both sets are measured, so an $X$ error on column $0$ flips a
vertical gauge and is detected while the new plaquette record is established.
From the following round the plaquettes are compared in time and the gauges may
be dropped. The minimum undetected $X$ string that flips $\bar Z$ therefore has
weight five, giving graphlike spacetime distance $d=m=5$, rather than the $d=1$
of an abrupt switch. The $X$-sector construction is dual: the $XX$ gauges remain
active while the row logical is extended across the resource columns. Stim's
shortest-graphlike-error diagnostic gives stitching-window distance $d=m$ in
both sectors for $m=3,5,7$. This stitching-window value is distinct from the
incoming end-to-end $X$-sector distance, which remains bounded by $n$.

The reverse $\mathrm{SC}(m\times m)\to\mathrm{BS}(m'\times m)$ interface uses
the same mechanism in the other sector, extending the transmitted $Z$ distance
from $m$ to $m'$. Both interfaces in the round trip therefore grow protection;
no size-reducing contraction is required.

\section{Deferred decoding and non-Clifford feedforward}
\label{app:nonclifford}

The equality between joint and phase-separated decoding used in the main text
holds for Clifford operations. Conjugating a Pauli through a $T$ gate leaves the
Pauli group, $T X T^\dagger=(X+Y)/\sqrt{2}$, so applying $T$ to a qubit carrying
an unresolved $X$-type frame produces a frame that cannot be tracked as a Pauli.
Equivalently, gate teleportation through a magic state leaves a byproduct $S^m$
conditioned on a logical measurement outcome $m$. Restoring a Pauli frame
requires applying $S^m$, and $m$ is known only after the corresponding logical
measurement has been decoded. This is the same adaptivity that separates
Pauli-basis cluster-state measurements, which require no feedforward, from
non-Pauli measurements whose bases depend on earlier outcomes
\cite{raussendorf2006,brown2020}.

Global error correction of the heterogeneous transmission-and-computation record
can still be deferred to the central node. What cannot be deferred is the
gadget-local feedforward decode that determines the conditional Clifford
correction at each non-Clifford gate. This operation is local in the global
history---it depends on the injection gadget rather than on the total round-trip
length---but remains a distance-$d$ logical-measurement decode and scales with the
local code distance. The client is therefore free of the global QEC decoder, not
of all local processing. A full non-Clifford round-trip study must determine
whether this feedforward can be resolved from the gadget's local syndrome alone
or whether its correctness couples to the upstream deferred record.

\end{document}